\documentclass[aps,prb,amsmath,twocolumn,amssymb,titlepage,superscriptaddress,showpacs]{revtex4-1}
\pdfoutput=1

\usepackage{soul}
\usepackage{epsfig}% Include figure files
\usepackage{graphicx}% Include figure files
\usepackage{lipsum}% http://ctan.org/pkg/lipsum
\usepackage{dcolumn}% Align table columns on decimal point
\usepackage{bm}% bold math
\usepackage[utf8]{inputenc}
\usepackage{graphicx} 
\usepackage{xcolor}
\usepackage{float}

\newcommand{\astcycl}{\mathrlap{\kern0.085em{\circlearrowright}}\ast}
\newcommand{\taucycl}{\mathrlap{\kern0.42em{\bullet}}\circlearrowright}

\setulcolor{red}
\setul{0.3ex}{0.4ex}
\begin{document}

\title{Electronic band structure in pristine and Sulfur-doped Ta$_2$NiSe$_5$  }
%\title{Evolution of the electronic band structure with Sulfur-doping in Ta$_2$NiSe$_5$  and its three-dimensionality revealed by angle-resolved photoemission}

\author{Tanusree Saha }
\altaffiliation{Corresponding author: tanusree.saha@student.ung.si}
\affiliation{Laboratory of Quantum Optics, University of Nova Gorica, 5001 Nova Gorica, Slovenia}

\author{Luca Petaccia}
\affiliation{Elettra Sincrotrone Trieste, Strada Statale 14 km 163.5, 34149 Trieste, Italy}

\author{Barbara Ressel}
\affiliation{Laboratory of Quantum Optics, University of Nova Gorica, 5001 Nova Gorica, Slovenia}
%\affiliation{Elettra Sincrotrone Trieste, Strada Statale 14 km 163.5, 34149 Trieste, Italy}

\author{Primo\v{z} Rebernik Ribi\v{c}}
%\affiliation{Laboratory of Quantum Optics, University of Nova Gorica, 5001 Nova Gorica, Slovenia.}
\affiliation{Elettra Sincrotrone Trieste, Strada Statale 14 km 163.5, 34149 Trieste, Italy}

\author{Giovanni Di Santo}
\affiliation{Elettra Sincrotrone Trieste, Strada Statale 14 km 163.5, 34149 Trieste, Italy}

\author{Wenjuan Zhao }
\affiliation{Elettra Sincrotrone Trieste, Strada Statale 14 km 163.5, 34149 Trieste, Italy}

\author{Giovanni De Ninno}
\affiliation{Laboratory of Quantum Optics, University of Nova Gorica, 5001 Nova Gorica, Slovenia}
\affiliation{Elettra Sincrotrone Trieste, Strada Statale 14 km 163.5, 34149 Trieste, Italy}

\begin{abstract}
We present an angle-resolved photoemission study of the electronic band structure of the excitonic insulator Ta$_2$NiSe$_5$, as well as its evolution upon Sulfur doping. Our experimental data show that while the excitonic insulating phase is still preserved at a Sulfur-doping level of 25$\%$, such phase is heavily suppressed when there is a substantial amount, $\sim$ 50$\%$, of S-doping at liquid nitrogen temperatures. Moreover, our photon energy-dependent measurements reveal a clear three dimensionality of the electronic structure, both in Ta$_2$NiSe$_5$ and Ta$_2$Ni(Se$_{1-x}$S$_x$)$_5$ ($x=0.25, 0.50$) compounds. This suggests a reduction of electrical and thermal conductivities, which might make these compounds less suitable for electronic transport applications.

\end{abstract}
\date{\today}

\maketitle
\section{Introduction}
The dimensionality and degree of anisotropy in electronic transport and optical properties of layered materials, such as, MoS$_2$\cite{mos2_1,mos2_2}, ReS$_2$\cite{res2_1} and other transition-metal dichalcogenides\cite{coleman}, make them potentially suitable for the fabrication of new electronic and optoelectronic devices. A recent addition to the family of such materials are the ternary chalcogenides, Ta$_2$NiX$_5$ (X $\rightarrow$ Se/S), which have attracted attention due to their layered crystalline structure and in-plane anisotropic properties\cite{tns1,tns2,tns3}. In both materials, the transition metal atoms and the chalcogen atoms are arranged forming a chain pattern in each layer and the layers are stacked by van der Waals forces\cite{structure}. While the ground state of Ta$_2$NiSe$_5$ is an excitonic insulator (EI)\cite{excitonic1,excitonic2,excitonic3}, Ta$_2$NiS$_5$ does not show any evidence of hosting an excitonic insulating phase and is a normal semiconductor in its ground state\cite{tnsulfur}. An excitonic insulator state is realised in materials where the valence and conduction bands are separated by a very small energy gap ($E_g>0$) or there is a small overlap in energies of the band edges ($E_g<0$). The attractive Coulomb force between electrons and holes in a narrow gap semiconductor or in a semimetal lead to the formation of bound electron-hole pairs, known as excitons\cite{EI_1,EI_2,EI_3,EI_4}. The formation of excitons in a system gives rise to an intriguing ground state, namely, the excitonic insulator\cite{EI_5}, which is characterised by a larger band gap that mirrors the exciton binding energy $E_b$.

The pre-requisite for the spontaneous formation of excitons is $\mid E_b \mid > \mid E_g \mid$ and the excitonic state will be most stable in a zero band gap material ($E_g=0$). Since $E_g$ is usually significantly larger than $E_b$ in semiconductors and insulators, materials with nearly zero energy gap or nearly zero energy overlap ($E_g\approx0$ such that a small $E_b$ can establish the EI ground state) are chosen as potential candidates in the search of excitonic insulators. The spontaneous formation of excitons is suppressed with increasing values of $\mid E_g \mid$, i.e., for more positive values of $E_g$ when $E_g>0$ (in a semiconductor) and for more negative values of $E_g$ when $E_g<0$ (in a semimetal), leading to a decrease in the values of the EI transition temperature $T_c$. As $\mid E_g \mid$ is increased such that it is comparable to or larger than the exciton binding energy, i.e., when $\mid E_g \mid\gtrsim E_b$, the excitonic phase becomes increasingly unstable against the semiconducting/semimetallic ground state\cite{EI_2,EI_4}. The energy gap $E_g$ can be controlled through chemical substitution or by applying physical pressure\cite{pressure1,pressure2}.

The EI phase has been identified in a number of materials, such as TiSe$_2$\cite{tise2_1,tise2_2}, Ta$_2$NiSe$_5$\cite{pressure2}, TmSe$_{1-x}$Te$_x$\cite{tst}, InAs/GaSb quantum wells\cite{inas}, etc. However, the phase transition in some of these materials is strongly influenced by lattice and spin degrees of freedom\cite{tise2_3,thermo}, for example, a charge density wave (CDW) or a band Jahn--Teller effect put forward for 1T-TiSe$_2$\cite{tise2_4,tise2_5}. In this respect, Ta$_2$NiSe$_5$ has emerged as a promising candidate for hosting a canonical excitonic insulator phase, excluding any alternative scenarios such as lattice distortion-mediated CDW mechanism leading to the phase transformation.  However, Ta$_2$NiSe$_5$ also undergoes an orthorhombic to monoclinic structural phase transition\cite{structure,disalvo} in addition to EI transition which raises a question on the dominant origin of the order parameter in this material. This triggered numerous experimental and theoretical\cite{theory1,theory2,theory3} studies that used either
equilibrium\cite{order1,order2,excitonic2,order3} or non-equilibrium \cite{order4,order5} approaches trying to resolve the
issue. The question on the dominant origin of the order parameter in
Ta$_2$NiSe$_5$ has also been addressed in one of our recent works\cite{prbsaha}. The scope of this paper is not to identify the origin of the phase transition but to investigate the effect of increasing $E_g$ on the electronic band structure of Ta$_2$NiSe$_5$ and the dimensionality in pure and doped compounds. The increase of $E_g$ is obtained by introducing Sulfur atoms at Selenium sites. Beyond a certain doping limit, $x$, the excitonic insulating phase in Ta$_2$Ni(Se$_{1-x}$S$_x$)$_5$ is completely suppressed\cite{pressure2}, a signature of which is the absence of a second-order phase transition in resisitivity measurements. As $x$ increases, changes in hybridization between Ni and Se/S valence band orbitals lead to a monotonic increase of the bandgap $E_g$ and, as consequence to a reduction of $T_c$. Application of physical pressure also results in the reduction of $T_c$ in pristine Ta$_2$NiSe$_5$\cite{pressure2}.

\begin{figure}[t]
\centering
\vspace{-4ex}
\includegraphics[width = 0.9\linewidth]{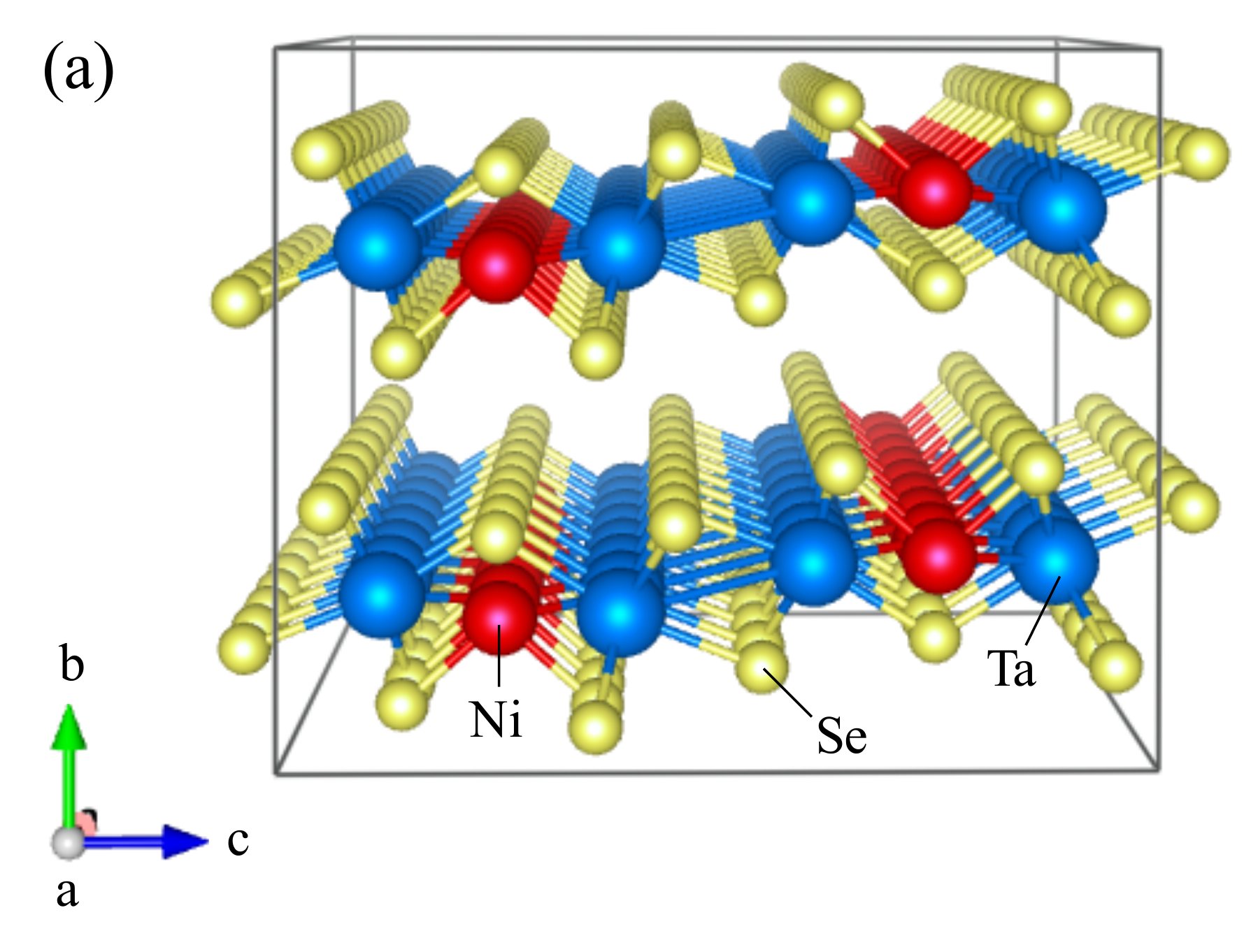}
\includegraphics[width = 1\linewidth]{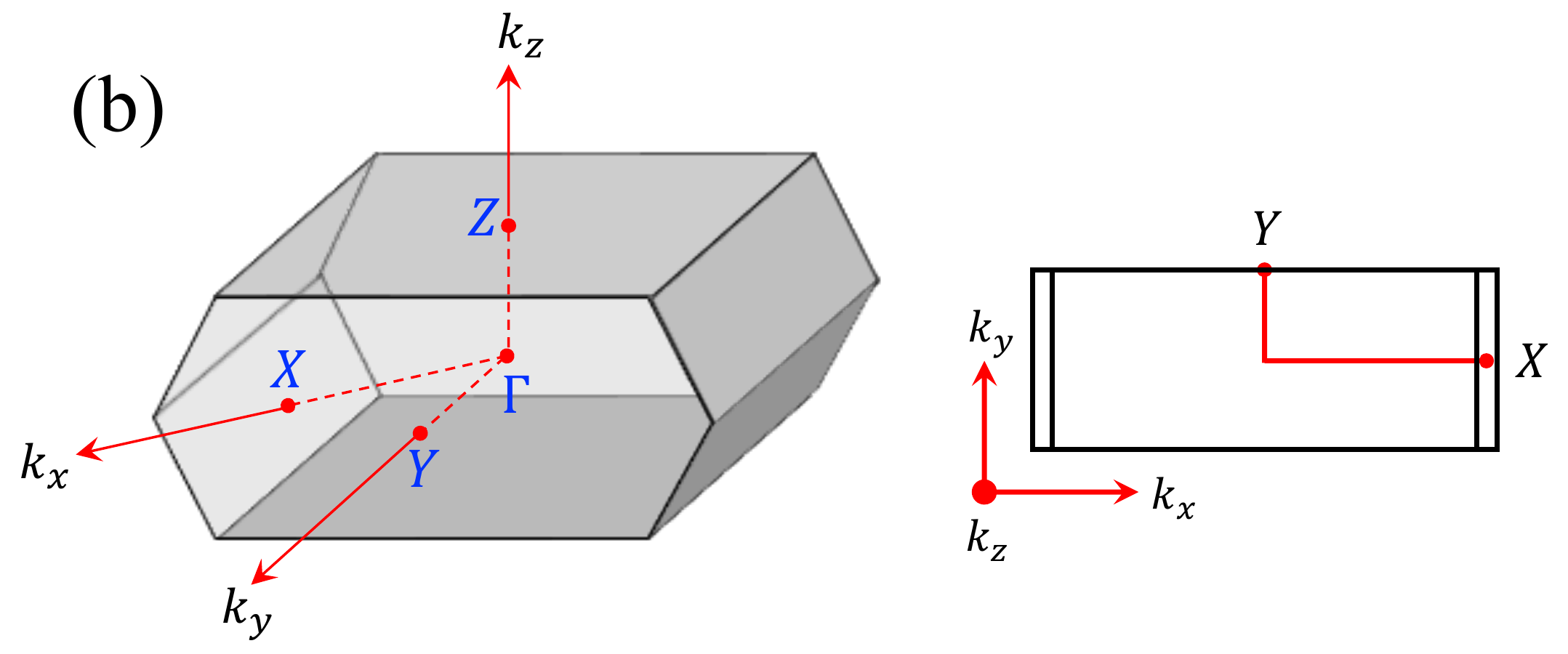}
\includegraphics[width = 1\linewidth]{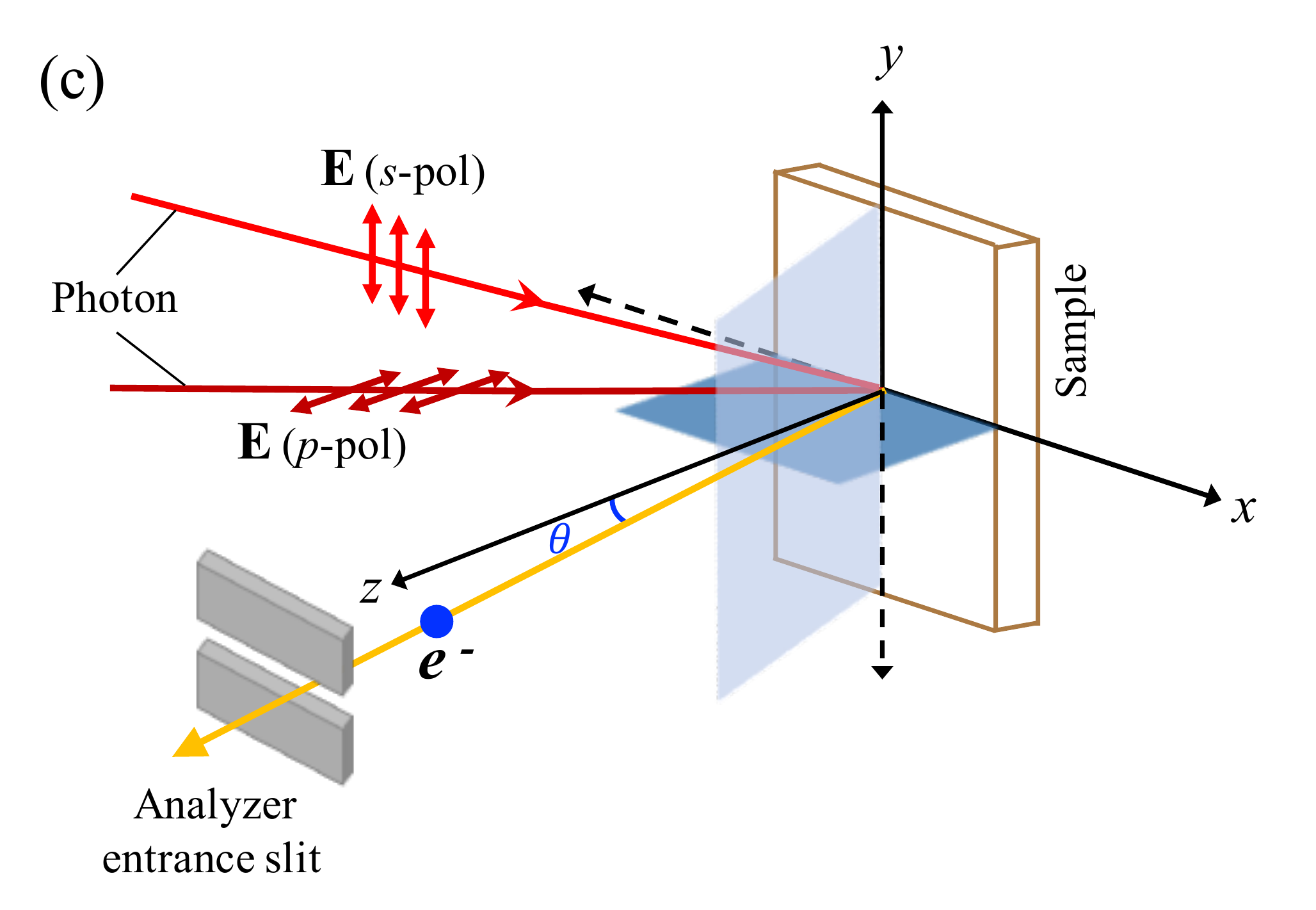}
\vspace{-2ex}
\caption{(a) Crystal structure of Ta$_2$NiSe$_5$ in the monoclinic phase where \textit{a}, \textit{b}, \textit{c} define the crystal axis system and tantalum, nickel and selenium atoms are represented by blue, red and yellow spheres, respectively. (b) (left) Bulk Brillouin zone (BZ) of a monoclinic lattice where $\Gamma,X,Y,Z$ are the high symmetry $k$-points, (right) Projection of the BZ on $k_xk_y$ plane. (c) Schematic of the experimental geometry showing the polarization of incident photons with respect to the sample. }
\label{figure0}
\vspace{-2ex}
\end{figure}   

\begin{figure}[t]
\centering
\vspace{-2ex}
\includegraphics[width = 1\linewidth]{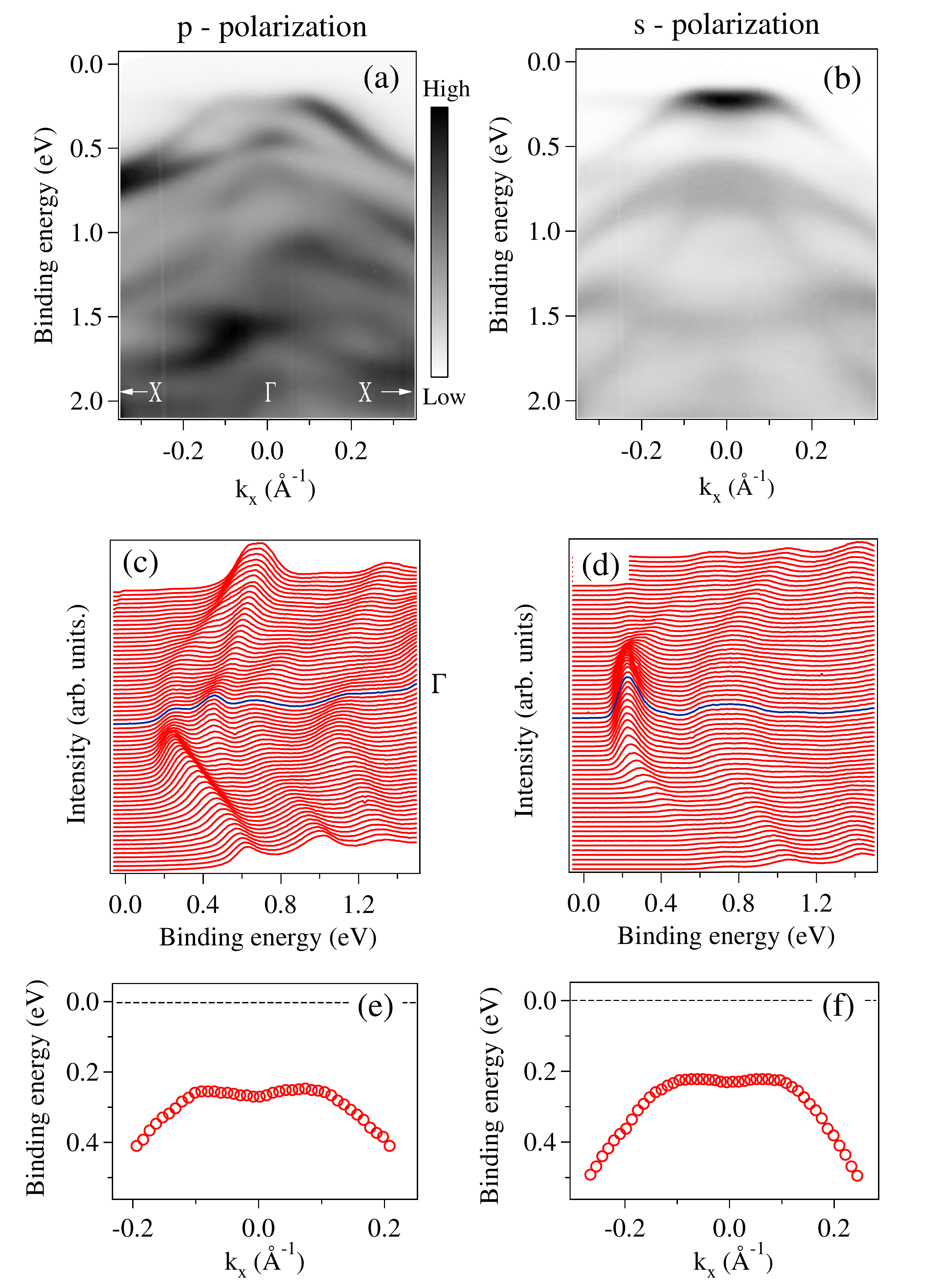}
\vspace{-2ex}
\caption{ARPES intensity maps of Ta$_2$NiSe$_5$ along $\Gamma-X$ direction acquired at T = 80 K using (a) p-polarized (horizontal) and (b) s-polarized (vertical) light of photon energy $h\nu$ = 19 eV. The extracted energy distribution curves (EDCs) stacked along $k_x$-axis for p-pol (c) and s-pol (d). The blue curve corresponds to the EDC at $\Gamma$ ($k_x=0$) in each case. A zoomed view of the valence band dispersion showing the band flatness near Fermi level, $E_F$, where the peak positions of the band obtained from the EDCs in (c) and (d) are plotted as a function of $k_x$ for both polarisations, (e) p-pol and (f) s-pol. The dashed line denotes the Fermi level, $E_F$.  }
\label{figure1}
\vspace{-2ex}
\end{figure}

Angle-resolved photoemission spectroscopy is an experimental technique for
probing the occupied energy bands in different solid state materials
and can be directly compared to band structure calculations. In the present work, we utilize polarization-dependent ARPES to investigate the effect of increasing Sulfur doping levels on the electronic band structure of Ta$_2$Ni(Se$_{1-x}$S$_x$)$_5$, notably the changes in $E$ vs. $k$ dispersion of the valence band, along different symmetry directions in $k$-space.  While extensive equilibrium and non-equilibrium studies have been done on pristine Ta$_2$NiSe$_5$, the $T_c$ vs. $x$ phase diagram\cite{pressure2} has not been explored through ARPES so far. Although the parent compounds are known to possess strongly two-dimensional properties of the electronic structures, which is consistent with the layered structure of Ta$_2$NiSe$_5$ and Ta$_2$NiS$_5$, our photon-energy dependent ARPES data reveal a possible three dimensionality of the electronic structure in Ta$_2$Ni(Se$_{1-x}$S$_x$)$_5$ compounds. 

\section{Experimental details}

Single crystals of Ta$_2$Ni(Se$_{1-x}$S$_x$)$_5$ (with $x$ = 0.0, $x$ = 0.25 and $x$ = 0.50) were purchased from HQ Graphene (http://www.hqgraphene.com). The
samples were synthesised by chemical vapour transport (CVT)\cite{cvt1,structure}. The ARPES experiments were carried out at the BaDElPh beamline\cite{badelph} of the Elettra synchrotron in Trieste, Italy. It offers both horizontal (p-pol) and vertical (s-pol) light polarizations, a maximum angular resolution of 0.1$^\circ$ and energy resolution of 5.4 meV. A photon energy range from $h\nu$ = 16 eV to $h\nu$ = 34 eV was used for this study and all the data presented in the paper were acquired at sample temperature T = 80 K. Prior to ARPES measurements, clean sample surfaces were obtained via cleaving in the direction perpendicular to the atomic planes. The samples were cleaved under UHV pressure better than 5 x 10$^{-10}$ mbar and the measurements were performed at a base pressure $<$ 1 x 10$^{-10}$ mbar.

%The ARPES experiments were performed at the BaDElPh\cite{badelph} beamline at {Elettra synchrotron} light source facility in Trieste (Italy), \ul{with a} SPECS Phoibos 150 electron analyzer. 

\section{Results and Discussion}

Ta$_2$NiSe$_5$, first reported in Ref.[8], has a layered crystalline structure, where each layer consists of double chains of tantalum (Ta) atoms with single chains of nickel (Ni) atoms in-between. The Ta-Ni-Ta trichains are parallel to the crystallographic $a$-axis and repeat themselves along the $c$-axis, exhibiting a quasi-one dimensional structure. Selenium atoms are tetrahedrally and octahedrally coordinated around the Ni atoms and Ta atoms, respectively. The layers are stacked along $b$-axis and are held together through van der Waals forces. The crystal structure of Ta$_2$NiSe$_5$ in its low temperature monoclinic phase is shown in Fig.~\ref{figure0}(a).  Band structure calculations have shown that the topmost valence band comprises Ni 3$d$ and Se 4$p$ orbitals while the conduction band bottom is composed of Ta 5$d$ orbitals\cite{excitonic3}. The high temperature orthorhombic phase of Ta$_2$NiSe$_5$ is characterised by a very small direct band gap while its low temperature monoclinic phase is an excitonic insulator, with a second-order phase transition occuring at $T_c\approx$ 326 K\cite{disalvo}. Excitons are formed by Ni 3$d$ - Se 4$p$ holes and Ta 5$d$ electrons with a flattened dispersion of the valence band top characterising the excitonic insulating state from an ARPES point of view. A schematic of the bulk Brillouin zone of Ta$_2$NiSe$_5$ for the monoclinic phase, along with its projection on the $k_xk_y$ plane is shown in Fig.~\ref{figure0}(b).

\begin{figure}[t]
\centering
\vspace{-2ex}
\includegraphics[width = 1\linewidth]{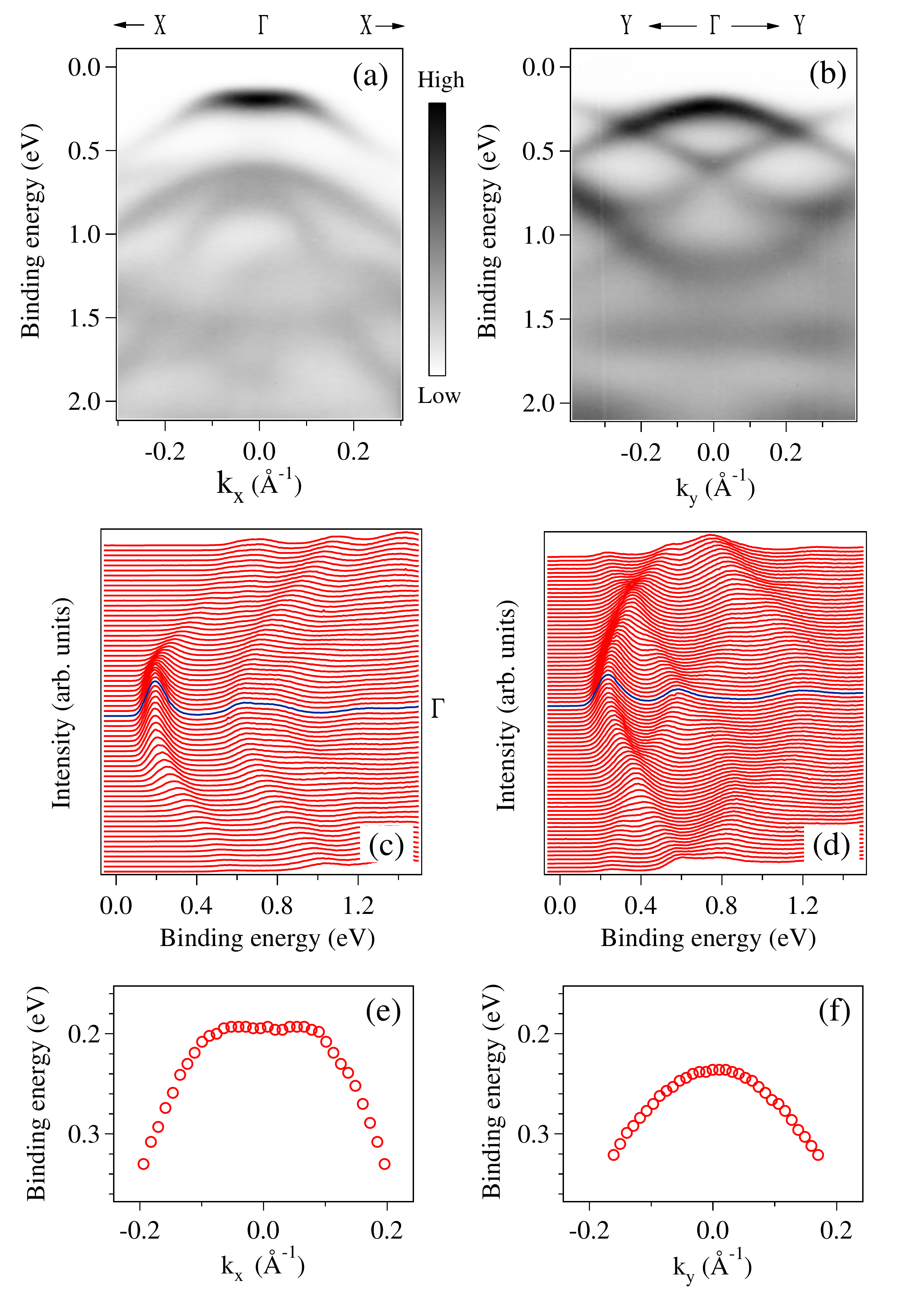}
\vspace{-2ex}
\caption{ARPES intensity maps of 25$\%$ S-doped Ta$_2$NiSe$_5$ along (a) $\Gamma-X$ and (b) $\Gamma-Y$ directions acquired at T = 80 K and photon energy $h\nu$ = 20 eV using $s$-polarized photons. The extracted energy distribution curves (EDCs) stacked along $k_x$-axis in (c) and $k_y$-axis in (d). The blue curve corresponds to the EDC at $\Gamma$ [$k_x=0$ in (a), $k_y=0$ in (b)]. A zoomed view of the valence band dispersion near Fermi level, $E_F$, where the peak positions of the EDCs in (c) and (d) are plotted as a function of $k_x$ and $k_y$ in (e) and (f), respectively. }
\label{figure2}
\vspace{-2ex}
\end{figure} 

\begin{figure*}[t]
\centering
\vspace{-2ex}
\includegraphics[width = 1\linewidth]{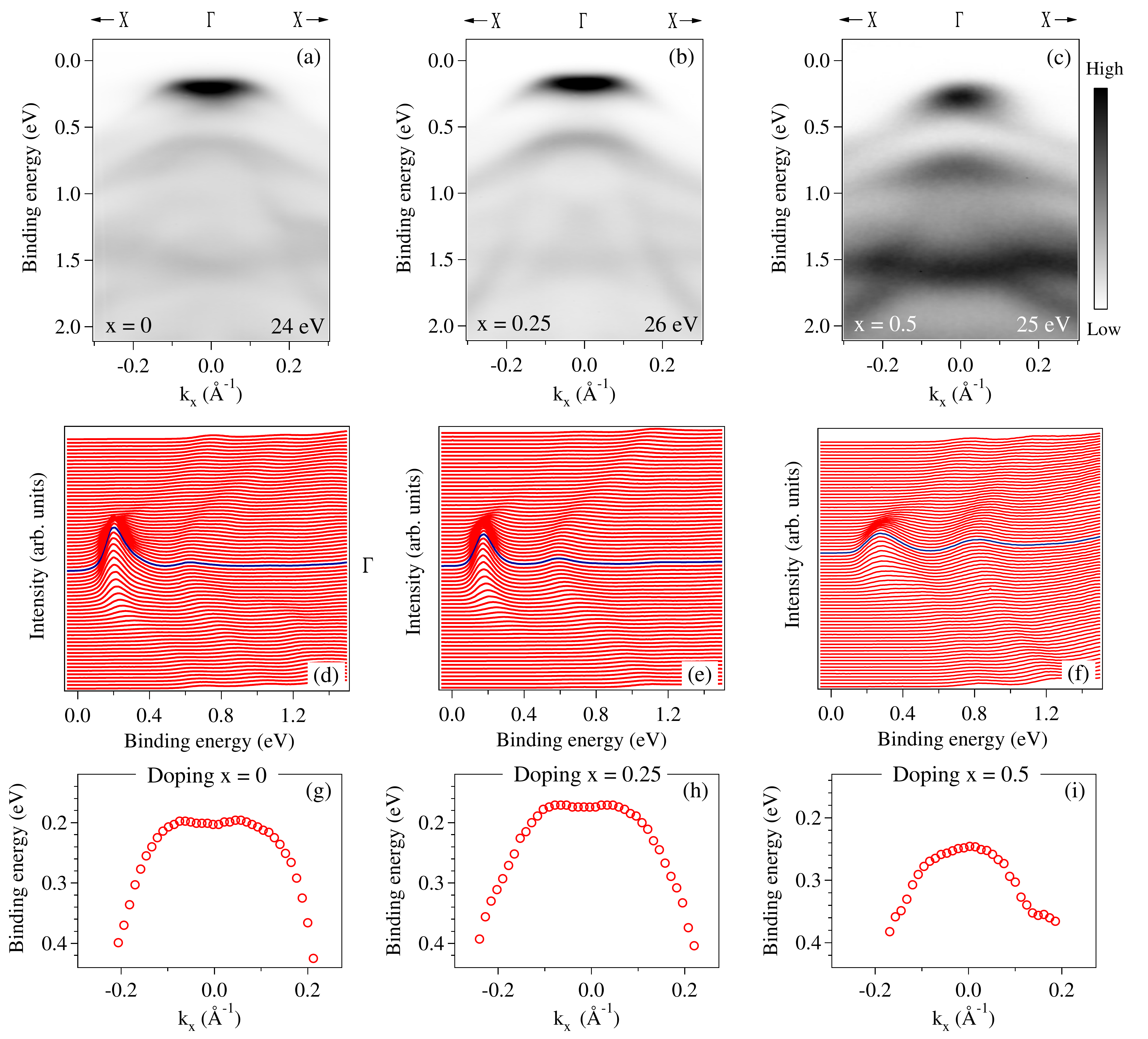}
\vspace{-2ex}
\caption{(a)-(c) ARPES intensity plots of Ta$_2$NiSe$_5$, Ta$_2$Ni(Se$_{0.75}$S$_{0.25}$)$_5$ and Ta$_2$Ni(Se$_{0.5}$S$_{0.5}$)$_5$, respectively, along $\Gamma-X$ direction acquired at photon energies $h\nu$ = 24 eV, 26 eV, 25 eV (s-polarization) respectively, and at T = 80 K. (d)-(f) Stacked energy distribution curves (EDCs) extracted from their respective ARPES plots in (a)-(c); the blue curve corresponds to the EDC at $\Gamma$ in each case. (g)-(i) Band dispersion of the top part of the valence band near $E_F$ for $x$ = 0, 0.25, 0.5, respectively, obtained from the peak positions of the corresponding EDCs. }
\label{figure3}
\vspace{-2ex}
\end{figure*} 

 The experimental geometry of the sample plane with respect to the analyser slit and direction of light polarization is schematically represented in Fig.~\ref{figure0}(c). Fig.~\ref{figure1} shows an overview of the band structure in Ta$_2$NiSe$_5$ for the low temperature phase, revealed by our ARPES measurements along $\Gamma-X$ direction, using light with different linear polarizations. The ARPES spectra in Figs.~\ref{figure1}(a)-(b) show that the top part of the valence band centered at $\Gamma$ ($k_{x}=0$) is characterised by a flat band dispersion\cite{excitonic1} which is the signature of an excitonic insulating phase. The stacked energy distribution curves (EDCs) along $k_{x}$ extracted from the (a)-(b) ARPES intensity plots are shown in Figs~\ref{figure1}(c)-(d), respectively, to give a better perspective of the band dispersions, especially the valence band below Fermi level, $E_F$ (referenced to 0). To give a clear picture of the characteristic band flatness close to $E_F$, the peak energy positions of the topmost valence band determined from the EDCs in (c) and (d) are plotted as a function of $k_{x}$ in Figs.~\ref{figure1}(e) and ~\ref{figure1}(f), respectively. We observe that the top valence band lies at $\sim$ 0.2 eV below $E_F$ and its dispersion around $\Gamma$ is hole-like with a flattened top, giving an overall shallow M-shaped dispersion. The parts away from $\Gamma$ and more towards $X$ are strongly dispersive. A characteristic M-shaped dispersion which becomes sharper on lowering the temperature has been previously reported from ARPES studies done using different light polarization geometries\cite{order3}. 

\begin{figure*}[t]
\centering
\vspace{-2ex}
\includegraphics[width = 1\linewidth]{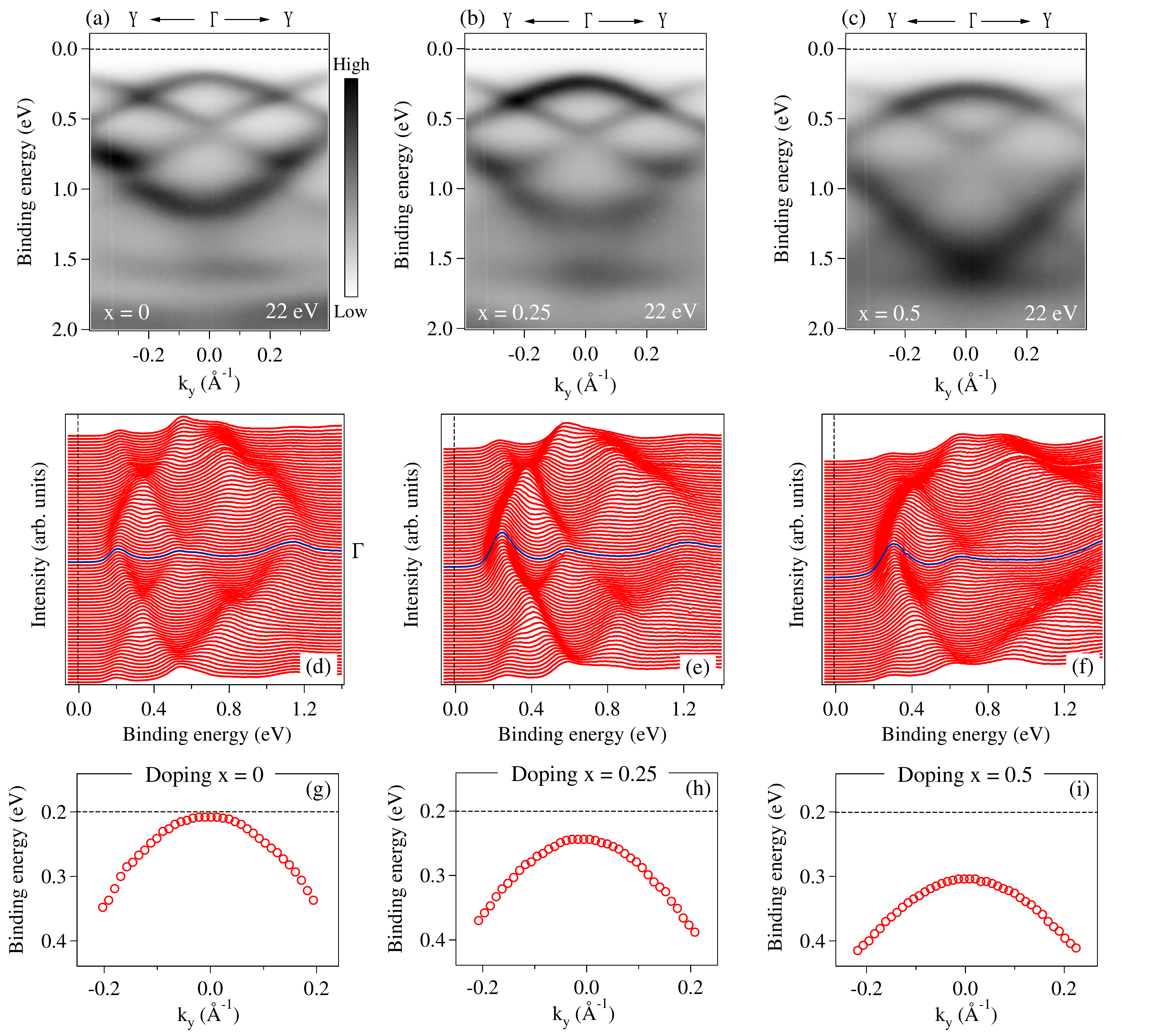}
\vspace{-2ex}
\caption{(a)-(c) ARPES intensity plots of Ta$_2$NiSe$_5$, Ta$_2$Ni(Se$_{0.75}$S$_{0.25}$ and Ta$_2$Ni(Se$_{0.5}$S$_{0.5}$)$_5$, respectively, along $\Gamma-Y$ direction acquired at photon energy $h\nu$ = 22 eV (s-polarization) and at T = 80 K. (d)-(f) Stacked energy distribution curves (EDCs) extracted from their respective ARPES plots in (a)-(c); the blue curve corresponds to the EDC at $\Gamma$ in each case. The dashed lines in (a)-(f) denote the Fermi level, $E_F$. (g)-(i) Band dispersion of the top part of the valence band near $E_F$ for $x$ = 0, 0.25, 0.5 respectively obtained from the peak position of the corresponding EDCs. The dashed line denotes the peak position at $\Gamma$ for $x$ = 0. }
\label{figure4}
\vspace{-2ex}
\end{figure*}

Moving to the polarization dependence of the ARPES spectra, we note from Figs.~\ref{figure1}(a) and ~\ref{figure1}(b) that the photoemission spectra  acquired using two different polarization geometries, namely, p-polarization (\textbf{E} lying in $xz$ plane in Fig.~\ref{figure0}(c)) and s-polarization (\textbf{E} along $y$-direction in Fig.~\ref{figure0}(c)), exhibit strong differences. The spectral weight of the flat top part of the valence band centered at $\Gamma$ is relatively strong for s-polarization. For the dispersive parts of the valence band which are away from $\Gamma$ and along the $\Gamma-X$ line, the spectral weight is stronger when probed with p-polarized light. This polarization-dependent contrast continues to exist at different photon energies explored and hence is attributed to the symmetry selection rules of the photoemission process\cite{harmanson,damascelli}. Similar symmetry contrasts are also observed in 25$\%$ and 50$\%$ S-doped Ta$_2$NiSe$_5$ (Supplementary Material, Section A). Such drastic changes in the spectral intensity of polarization-dependent ARPES data can help us identify the mirror symmetries associated with different atomic orbitals. From now on, we will only show the data acquired using s-polarized light due to the better visibility of the top part of the
valence band, which is relevant for the excitonic insulator phase. 
%ARPES with linearly polarized light can be used to identify the mirror symmetry of the electron orbitals in the occupied bands\cite{symmetry}. 

Although Ta$_2$NiSe$_5$ is believed to form a quasi-one-dimensional structure, the hopping of mobile carriers along $c$-axis gives rise to a two-dimensional (2D) character of the electronic band structure.   There exists an anisotropy in the band dispersions of Ta$_2$NiSe$_5$\cite{aniso,order3,symmetry} and Ta$_2$NiS$_5$\cite{tnsulfur} along $\Gamma-X$ and $\Gamma-Y$ directions. Our ARPES data reveal similar anisotropic properties of the electronic structure in Sulfur-doped Ta$_2$NiSe$_5$ as well. The ARPES intensity maps along $\Gamma-X$ and $\Gamma-Y$ (refer to Fig.~\ref{figure0}(b) for the $k$-space symmetry points) of Ta$_2$Ni(Se$_{1-x}$S$_x$)$_5$ for $x=0.25$ are shown in Figs.~\ref{figure2}(a) and \ref{figure2}(b), respectively. We observe that the dispersion around $\Gamma$ point along $k_x$-axis ($\Gamma-X$ direction) is characterised by a flat feature of the valence band top while the dispersion of the valence band at $\Gamma$ along $k_y$-axis ($\Gamma-Y$ direction) does not exhibit the hybridized flattened band dispersion. The periodicity of the band dispersion along $k_y$ in \ref{figure2}(b) could be easily captured with the photon energy range used in this study, due to its small periodicity along $\Gamma-Y$ direction. For better clarity of the band dispersions along both directions, the corresponding stacked EDC plots are displayed in Figs.~\ref{figure2}(c) and \ref{figure2}(d). A zoomed view of the dispersion for the top part of the valence band centered at $\Gamma$ along $k_x$ and $k_y$ directions is presented in Figs.~\ref{figure2}(e) and \ref{figure2}(f), respectively. The dispersions have been obtained from the peak energy positions of the corresponding EDCs (\ref{figure2}(c) and \ref{figure2}(d)) for the valence band top below $E_F$. The flat band feature around $\Gamma$ along $k_x $ is clearly seen while the dispersion of the valence band top along $k_y$ seems to be parabolic. For completeness, the anisotropic dispersions for pristine Ta$_2$NiSe$_5$ and Ta$_2$Ni(Se$_{1-x}$S$_x$)$_5$ ($x=0.5$) are presented in the Supplementary Material (Section B). Because the bands also show a dispersion in the (in-plane) direction perpendicular to the chains, the quasi-1D structure often assumed for this family of compounds does not fully describe the electronic structure in these materials.

\begin{figure}[t]
\centering
\vspace{-2ex}
\includegraphics[width = 1\linewidth]{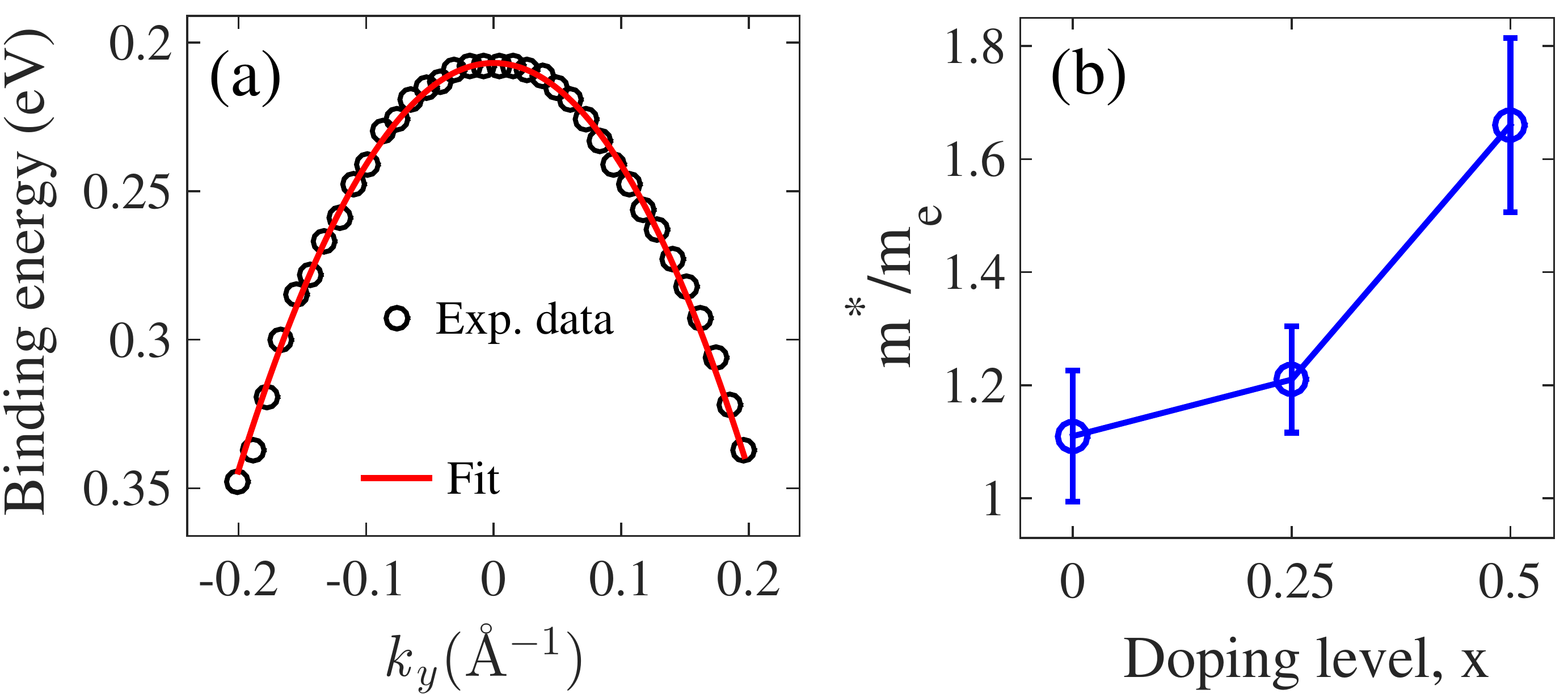}
\vspace{-4ex}
\caption{(a) Fit (red curve) to the $E$ vs. $k$ dispersion (black circles) around $\Gamma$ in pristine Ta$_2$NiSe$_5$. (b) Effective mass $m^*$ (in units of electron rest mass, $m_e$) at different doping levels, $x$.}
%\caption{Comparison of the valence band dispersions centered at $\Gamma$ along $\Gamma-Y$ direction between Ta$_2$NiSe$_5$, Ta$_2$Ni(Se$_{0.75}$S$_{0.25}$)$_5$ and Ta$_2$Ni(Se$_{0.5}$S$_{0.5}$)$_5$ on the same binding energy scale.}
\label{figure4a}
\vspace{-4ex}
\end{figure}

\begin{figure*}[t]
\centering
\vspace{-2ex}
\includegraphics[width = 1\linewidth]{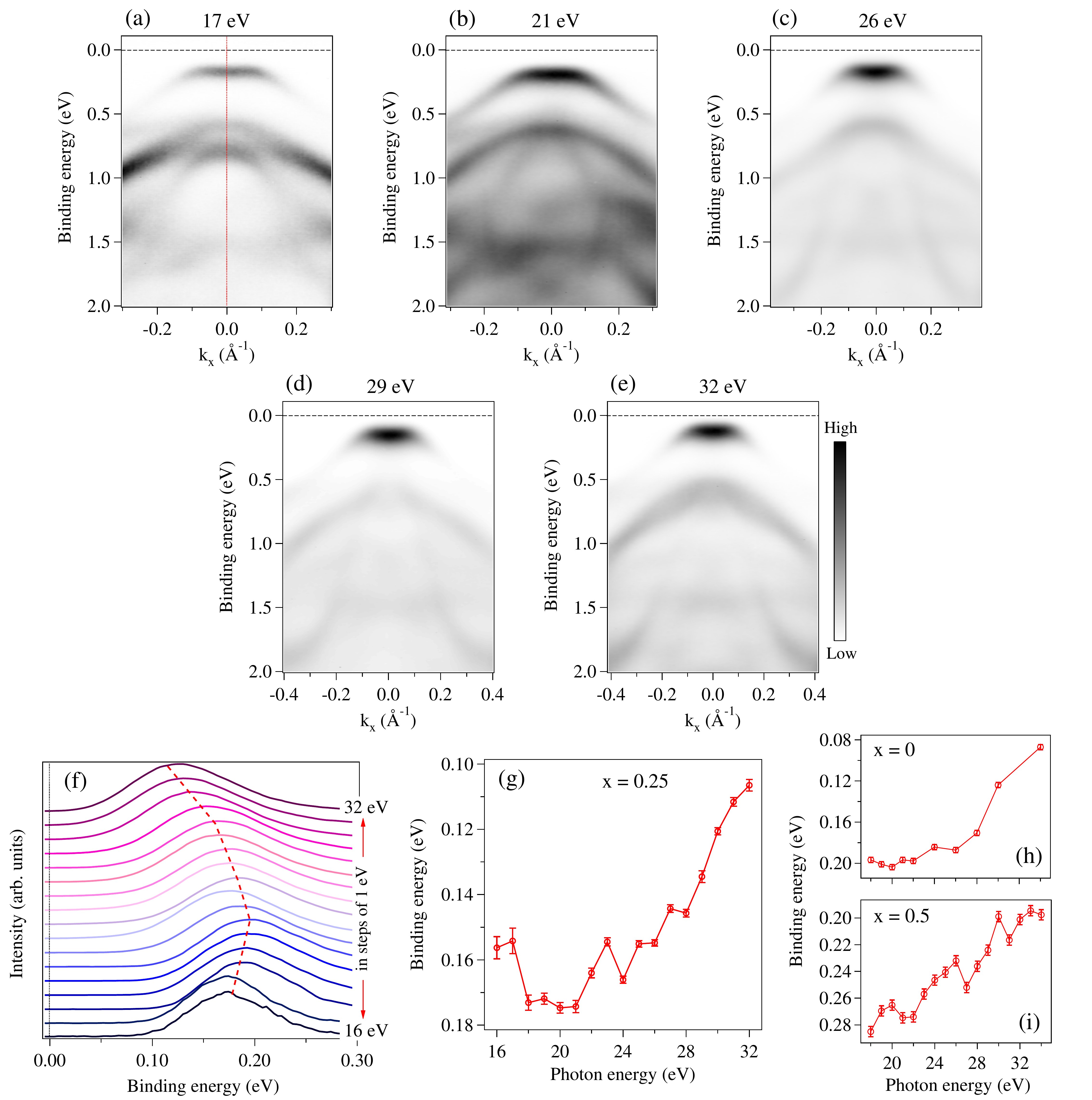}
\vspace{-2ex}
\caption{(a)-(e) ARPES intensity plots of Ta$_2$Ni(Se$_{0.75}$S$_{0.25}$)$_5$ (T = 80 K) along $\Gamma-X$ direction for s-polarized light at photon energies, $h\nu=$17 eV, 21 eV, 26 eV, 29 eV, 32 eV respectively. The $\Gamma$-point is denoted by the red dotted line in (a) and the dashed lines in (a)-(e) at 0 eV binding energy denote $E_F$. (f) EDCs at $\Gamma$ for different photon energies covering a range from $h\nu=$16 eV to 32 eV with a step size of 1 eV. The red dashed line is a guide to the eye following the peak position of the valence band with increasing photon energy. (g) Peak position of the valence band obtained by fitting the EDCs with a Lorentzian-Gaussian line shape at $\Gamma$ as a function of photon energy. (h), (i) Dependence of the valence band peak position on photon energy in pristine and 50$\%$ S-doped Ta$_2$NiSe$_5$, respectively. }
\label{figure6}
\vspace{-2ex}
\end{figure*}

\begin{figure*}[t]
\centering
\vspace{-2ex}
\includegraphics[width = 1\linewidth]{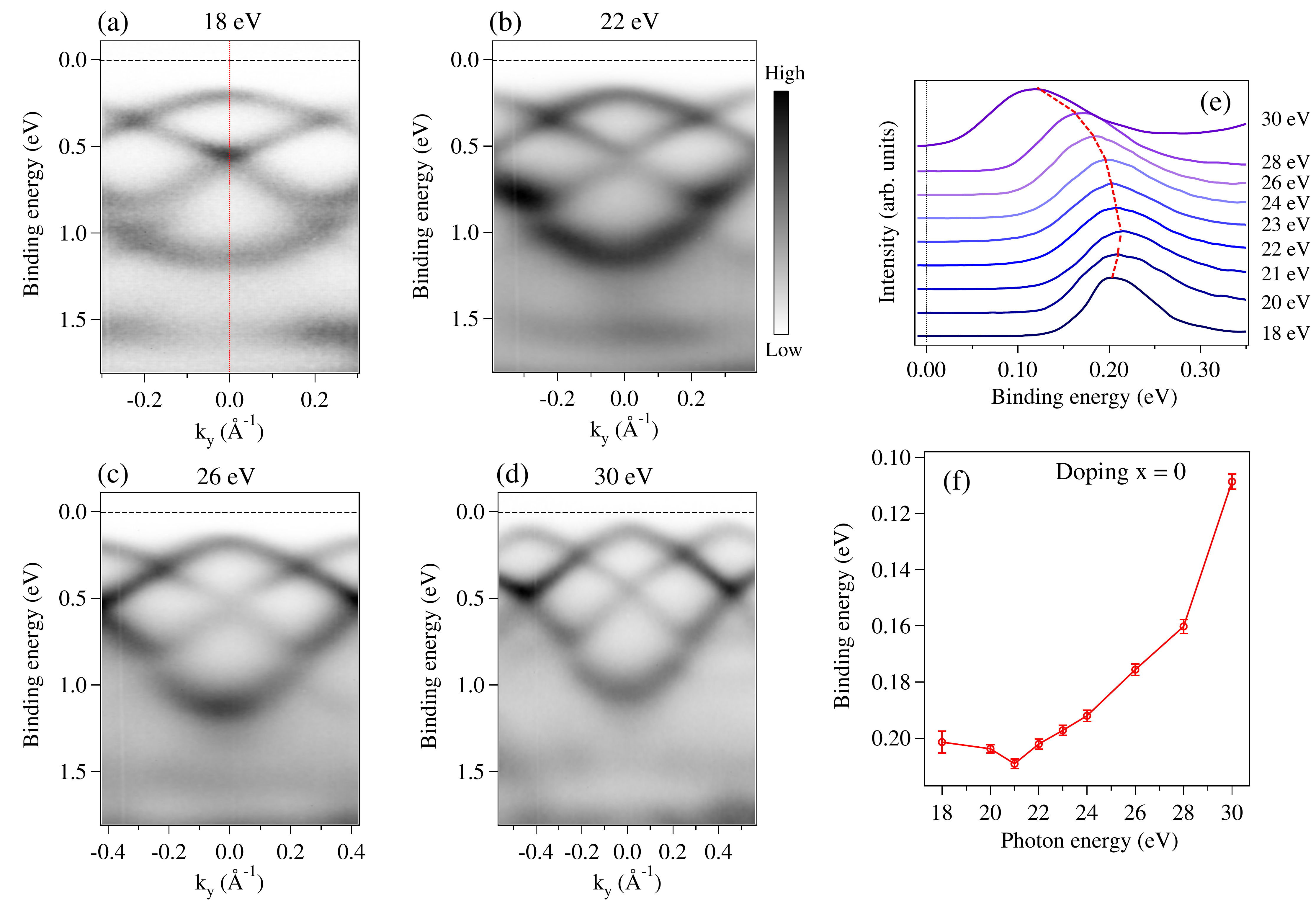}
\vspace{-2ex}
\caption{(a)-(d) ARPES intensity plots of pristine Ta$_2$NiSe$_5$ (T = 80 K) along $\Gamma-Y$ direction for s-polarized light at photon energies, $h\nu=$18 eV, 22 eV, 26 eV, 30 eV respectively. The $\Gamma$-point is denoted by the red dotted line in (a) and  the dashed lines in (a)-(d) at 0 eV binding energy denote $E_F$. (e) EDCs at $\Gamma$ for various photon energies. The red dashed line is a guide to the eye following the peak position of the valence band with increasing photon energy. (f) Peak position of the valence band obtained by fitting the EDCs at $\Gamma$ as a function of photon energy. }
\label{figure8}
\vspace{-2ex}
\end{figure*}

We now study the effect of Sulphur doping on the electronic structure of Ta$_2$NiSe$_5$, i.e., the band dispersions in Ta$_2$Ni(Se$_{1-x}$S$_x$)$_5$, with increasing values of $x$. Fig.~\ref{figure3} shows the evolution of the band dispersion along $\Gamma-X$ direction in Ta$_2$Ni(Se$_{1-x}$S$_x$)$_5$ with $x$ changing from 0 to 0.5. We observe that for Sulfur doping levels of $x=0, 0.25$ and 0.5, the band structure retains the main dispersive features down to approximately 2 eV below $E_F$, except for the top-part of the valence band at $\Gamma$. The relative changes in intensity are due to different photoionization cross sections at different photon energies. What is interesting to note here is that while the band flatness with the shallow M-shape is present for pristine Ta$_2$NiSe$_5$ (0$\%$ S-doping in Figs.~\ref{figure3}(a), (d)) and remains unaffected under 25$\%$ Sulfur doping (see Figs.~\ref{figure3}(b), (e)), it is significantly distorted for a substantial S-doping level of 50$\%$ (see Figs.~\ref{figure3}(c), (f)). Since the flattened dispersion of the valence band top is a fingerprint of the excitonic insulator phase, the observed deviation is indicative of the suppression of an excitonic insulator phase under heavy S-doping. For better clarity, the valence band dispersions around $\Gamma$ extracted from their respective EDC plots are displayed in Figs.~\ref{figure3}(g)--(i). The flat band distortion can be understood considering the size of the band gap $E_g$ in the high temperature phase of Ta$_2$Ni(Se$_{1-x}$S$_x$)$_5$. Indeed, with increasing levels of Sulfur doping, the band gap $E_g$ monotonically increases, thereby, suppressing the spontaneous formation of excitons. Hence, the excitonic insulating state becomes unstable against the semiconducting ground state. This emphasizes that although an excitonic insulator phase in Ta$_2$NiSe$_5$ can be realized with 25$\%$ of S-doping at 80 K, the excitonic ground state is destroyed when the doping level is $\approx$ 50$\%$. Electrical transport studies have revealed that a similar S-doping level (i.e., $x\geq0.5$) does not allow to observe the excitonic phase transition, down to 2K.

%a similar S-doping level of $x \geq 0.5$ in Ta$_2$Ni(Se$_{1-x}$S$_x$)$_5$ at which no excitonic phase transition was observed at temperatures down to 2 K\cite{pressure2} and the ground state remained semiconducting. 

We now turn to the evolution of the band structure along $\Gamma-Y$ direction as a function of S-doping. Figs.~\ref{figure4}(a)-(c) shows the band structure of Ta$_2$Ni(Se$_{1-x}$S$_x$)$_5$ for $x$ = 0, 0.25, 0.5 and the corresponding stacked EDCs are plotted in Figs.~\ref{figure4}(d)-(f) for better visualization of the band dispersions. As already shown in Figs.~\ref{figure2}(b), (d) and (f), the valence band near $E_F$ is characterised by a hole-like parabolic dispersion along $k_y$ centred at $\Gamma$-point.  From the ARPES intensity maps [Figs.~\ref{figure4}(a)-(c)], we see that the band dispersions along $k_y$-direction have subtle changes in the binding energies of different states for 25$\%$ S-doping [Fig.\ref{figure4}(e)] but pronounced changes for 50$\%$ S-doping [Fig.~\ref{figure4}(f)], when compared to the energy states in pristine Ta$_2$NiSe$_5$ [Fig.~\ref{figure4}(d)]. The dispersion of the valence band around $\Gamma$ near $E_F$ for the three doping levels are compared in Figs.~\ref{figure4}(g)-(i). We notice that by increasing the amount of S-doping, the parabolic dispersion centered at $\Gamma$ shifts to higher binding energies resulting in an overall enhancement of the energy gap along $\Gamma-Y$ direction. As stated before, the valence band is composed of Ni 3$d$ and Se 4$p$ (S 3$p$) orbitals. Therefore, increasing substitution by Sulfur atoms at the Se atom sites brings the states at the top of the valence band to higher binding energies. Another interesting fact that can be noted from the band dispersions is the change in effective mass ratio, $m^*/m_e$ (where, $m_e$ is the rest mass of an electron), in Ta$_2$Ni(Se$_{1-x}$S$_x$)$_5$ at different S-doping levels. The valence band dispersions along $\Gamma-Y$ have been fitted with quadratic functions, $E_0-\hbar^2k^2/2m^*$ ($h$ is the Planck's constant), with two fitting parameters $E_0$ and $m^*$. The experimental data in Figs.~\ref{figure4}(g)-(i) are well reproduced by the fits, an example of which is shown in Fig.~\ref{figure4a}(a). The values of $m^*$ obtained from the fits are plotted in units of $m_e$ as a function of S-doping level in Fig.~\ref{figure4a}(b). We observe that while there is a small difference in the values of $m^*/m_e$ between $x=0$ and $x=0.25$, it is comparatively larger for $x=0.5$. This is expected due to stronger carrier localization with increasing S-doping levels which arises from hybridized Ni 3$d$ and S 3$p$ orbitals, leading to higher effective masses. Such a result is consistent with our observations in Figs.~\ref{figure4}(g)-(i), where the energy states move towards higher binding energies as S-doping is increased. 
 
%along $\Gamma-Y$ are plotted on \ul{the same relative scale} for the three compounds in Fig.~\ref{figure5}. We observe that while the $E-k$ dispersion for $x=0$ and $x=0.25$ displays a negligible difference in the effective mass according to the relation $E\sim-\hbar^2k^2/2m^*$ (where $\hbar=h/2\pi$, $h$ is the Planck's constant), the dispersion for $x=0.5$ shows a larger value for $m^*$. This is expected due to stronger carrier localization with increasing S-doping levels which arises from hybridized Ni 3$d$ and S 3$p$ orbitals, leading to higher effective masses. Such a result is consistent with our observations in Figs.~\ref{figure4}(g)-(i), where the energy states move towards higher binding energies as S-doping is increased.  

Next, we investigate whether the band dispersion in Ta$_2$Ni(Se$_{1-x}$S$_x$)$_5$ exhibits a three-dimensional character. The three dimensionality of the electronic structure can be determined by probing the out-of-plane $\Gamma-Z$ direction which corresponds to $k_\perp$ in our experimental geometry. A photon energy dependent ARPES study is required to obtain such an information. Figures~\ref{figure6}(a)-(e) shows the $E$ vs. $k$ maps of Ta$_2$Ni(Se$_{1-x}$S$_x$)$_5$, $x=0.25$, along $\Gamma-X$ direction acquired at different photon energies. Due to low photoionization cross section at photon energies below 15 eV and above 35 eV, a photon energy range from 16 eV to 34 eV was chosen. Looking at ARPES intensity maps, one can notice a photon energy-dependent cross sectional contrast, as well as a considerable variation of the binding energies of different bands, especially for bands lying between 0.5 eV and 1 eV below $E_F$ and the flat band near $E_F$. We will mainly concentrate on the energy modifications of the `flat band' with variations in photon energy. The EDCs at $\Gamma$-point [along the red dotted line in Fig.~\ref{figure6}(a)] within 0.3 eV below $E_F$, capturing only the spectral intensity of the flat band, are plotted for different photon energies in Fig.~\ref{figure6}(f). From the ARPES images and the EDC plots, the obvious trend that we observe is the shift of the flat band towards $E_F$ on increasing the photon energy. This is clearly seen if one notes the energy separation between 0 eV binding energy (or $E_F$, shown by black dashed line) and the flat band below it in Figs.~\ref{figure6}(a)-(e). The dispersion of the flat band along $k_\perp$ can be visualized from the red dashed line denoting the peak position of the EDCs at different photon energies in Fig.~\ref{figure6}(f). The peak position of the flat band relative to $E_F$ at different photon energies has been obtained by fitting each of the EDCs with a Gaussian-Lorentzian distribution function after Shirley background subtraction. Fig.~\ref{figure6}(g) shows that the peak position of the flat band at $\Gamma$ moves towards $E_F$ with increasing photon energy, with a dip at $h\nu \approx20$ eV. A decrease in the binding energy of the flat band in Ta$_2$NiSe$_5$ and the distorted flat band in Ta$_2$Ni(Se$_{0.5}$S$_{0.5}$)$_5$ at $\Gamma$-point ($\Gamma-X$ direction) with increasing photon energy are shown in Figs.~\ref{figure6}(h) and (i), respectively. There are no pronounced qualitative differences in the trend of binding energy vs photon energy curves amongst the three compounds (Supplementary Material, Section C). For a perfectly two-dimensional electronic structure often used for Ta$_2$NiSe$_5$ and Ta$_2$NiS$_5$, the band dispersions should be independent of $k_z$ ($k_\perp$ in our experimental geometry) but the above observations clearly show a dispersion along $k_z$, i.e., along $\Gamma-Z$ direction in Fig.~\ref{figure0}(b). This demonstrates the three-dimensionality of electronic band structure in this family of compounds.  

For completeness, we also present the photon energy dependence of the band structure along $\Gamma-Y$ direction. Figs.~\ref{figure8}(a)-(d) show the ARPES intensity plots at different photon energies for pristine Ta$_2$NiSe$_5$. Again, an overall energy shift of the band structure towards $E_F$ (denoted by black dashed lines in the 2D-plots) is observed with increasing photon energies. The EDCs containing only the top-part of the valence band at $\Gamma$ are plotted for various photon energies in Fig.~\ref{figure8}(e). A similar trend in the shift of valence band peak position as that in Fig.~\ref{figure6}(f) can be noted here (marked by the red dashed line). The photon energy-dependent peak positions are plotted in Fig.~\ref{figure8}(f), where a decrease in binding energy is observed, thereby demonstrating a clear dispersion along $\Gamma-Z$ direction. The above results of the photon energy dependent ARPES measurements in all the three compounds do emphasize that inspite of the well known layered structure characteristing the Ta$_2$Ni(Se$_{1-x}$S$_x$)$_5$ compounds, the electronic structure is not strictly two-dimensional.

\section{Conclusion}
To summarize, we investigated the evolution of the band structure in Ta$_2$Ni(Se$_{1-x}$S$_x$)$_5$, as a function of Sulfur concentration $x$ and its three-dimensionality using ARPES. The anisotropy of the in-plane band dispersions along $\Gamma-X$ and $\Gamma-Y$ directions in the doped compounds is clearly revealed. A substantial amount of Sulfur doping $\sim$ 50$\%$ is able to suppress the excitonic ground state in Ta$_2$Ni(Se$_{1-x}$S$_x$)$_5$. This can be claimed due to the observed deviation of the dispersion of valence band top from its flattened $E$ vs. $k$ feature along $\Gamma-X$ at $x=0.5$ but not at $x=0.25$. The
suppression of the excitonic insulator state is followed by a
pronounced increase in the effective mass $m^*$ for the highest doping
level, indicating stronger localization of the charge carriers, which is in agreement with the weakening of hybridization betweeen Ni 3$d$ and Se 4$p$ (S 3$p$) orbitals that also leads to larger values of $E_g$. Our photon-energy dependent ARPES data show a gradual decrease in the binding energy of states near Fermi level on increasing the photon energy (increasing $k_z$). This emphasizes that the two-dimensional
picture often used for Ta$_2$NiSe$_5$ (and Ta$_2$NiS$_5$, not studied in this work) does not fully represent the entire electronic structure and that three-dimensionality must be taken into account when studying Ta$_2$Ni(Se$_{1-x}$S$_x$)$_5$ compounds. Dimensionality of the underlying electronic structure in a material is an important parameter while choosing potential candidates for electronic applications. Low-dimensional systems have surpassed their bulk counterparts in this field due to their high electrical and thermal conductivities. Our results therefore suggest that the three-dimensional nature of Ta$_2$Ni(Se$_{1-x}$S$_x$)$_5$ needs to be taken into account when considering these compounds for applications in the field of electronics and optoelectronics.

%Therefore, our results indicate that these materials might not have high potential for applications in the field of electronics and optoelectronics.

\acknowledgements
This project has received funding from the European Union’s Horizon 2020 research and innovation programme under grant agreement No 654360 NFFA-Europe. The authors acknowledge Elettra Sincrotone Trieste for providing access to its synchrotron radiation facilities and to the BaDElPh beamline that contributed to the results presented in this work. We thank L. Sancin for technical assistance during experiments and J. Mravlje and F. Galdenzi for fruitful discussions during the beamtime.

\end{document}

\title{Supplementary Material for: 
\\
Electronic band structure in pristine and Sulfur-doped Ta$_2$NiSe$_5$ }

\author{Tanusree Saha }
\altaffiliation{Corresponding author: tanusree.saha@student.ung.si}
\affiliation{Laboratory of Quantum Optics, University of Nova Gorica, 5001 Nova Gorica, Slovenia.}

\author{Luca Petaccia}
\affiliation{Elettra Sincrotrone Trieste, Strada Statale 14 km 163.5, 34149 Trieste, Italy}

\author{Barbara Ressel}
\affiliation{Laboratory of Quantum Optics, University of Nova Gorica, 5001 Nova Gorica, Slovenia.}
%\affiliation{Elettra Sincrotrone Trieste, Strada Statale 14 km 163.5, 34149 Trieste, Italy}

\author{Primo\v{z} Rebernik Ribi\v{c}}
%\affiliation{Laboratory of Quantum Optics, University of Nova Gorica, 5001 Nova Gorica, Slovenia.}
\affiliation{Elettra Sincrotrone Trieste, Strada Statale 14 km 163.5, 34149 Trieste, Italy}

\author{Giovanni Di Santo}
\affiliation{Elettra Sincrotrone Trieste, Strada Statale 14 km 163.5, 34149 Trieste, Italy}

\author{Wenjuan Zhao }
\affiliation{Elettra Sincrotrone Trieste, Strada Statale 14 km 163.5, 34149 Trieste, Italy}

\author{Giovanni De Ninno}
\affiliation{Laboratory of Quantum Optics, University of Nova Gorica, 5001 Nova Gorica, Slovenia.}
\affiliation{Elettra Sincrotrone Trieste, Strada Statale 14 km 163.5, 34149 Trieste, Italy}

%\begin{abstract}
%We investigate the non-equilibrium electronic structure and characteristic time scales in a candidate excitonic insulator, Ta$_2$NiSe$_5$, using time- and angle-resolved photoemission spectroscopy with a temporal resolution of 50 fs. Following a strong photoexcitation, the band gap closes transiently within 100 fs, i.e., on a time scale faster than the typical lattice vibrational period. Furthermore, we find that the characteristic time associated with the rise of the photoemission intensity above the Fermi energy decreases with increasing excitation strength, while the relaxation time of the electron population towards equilibrium shows an opposite behaviour. We argue that these experimental observations can be consistently explained by an excitonic origin of the band gap in the material. The excitonic picture is supported by microscopic calculations based on the non-equilibrium Green's function formalism for an interacting two-band system. We interpret the speedup of the rise time with fluence in terms of an enhanced scattering probability between photo-excited electrons and excitons, leading to an initially faster decay of the order parameter. We show that the inclusion of electron-phonon coupling at a semi-classical level changes only the quantitative aspects of the proposed dynamics, while the qualitative features remain the same. The experimental observations and microscopic calculations allow us to develop a simple and intuitive phenomenological model that captures the main dynamics after photoexcitation in Ta$_2$NiSe$_5$.
%\end{abstract}
\date{\today}

\maketitle
%\section{Introduction}

%\section{Supplementary Material}
\beginsupplement

\tableofcontents 

\clearpage

\subsection{Polarization dependent ARPES data in Ta$_2$Ni(Se$_{0.75}$S$_{0.25}$)$_5$ and Ta$_2$Ni(Se$_{0.5}$S$_{0.5}$)$_5$}

\begin{figure*}[h]
\centering
\vspace{4ex}
\includegraphics[width = 0.6\linewidth]{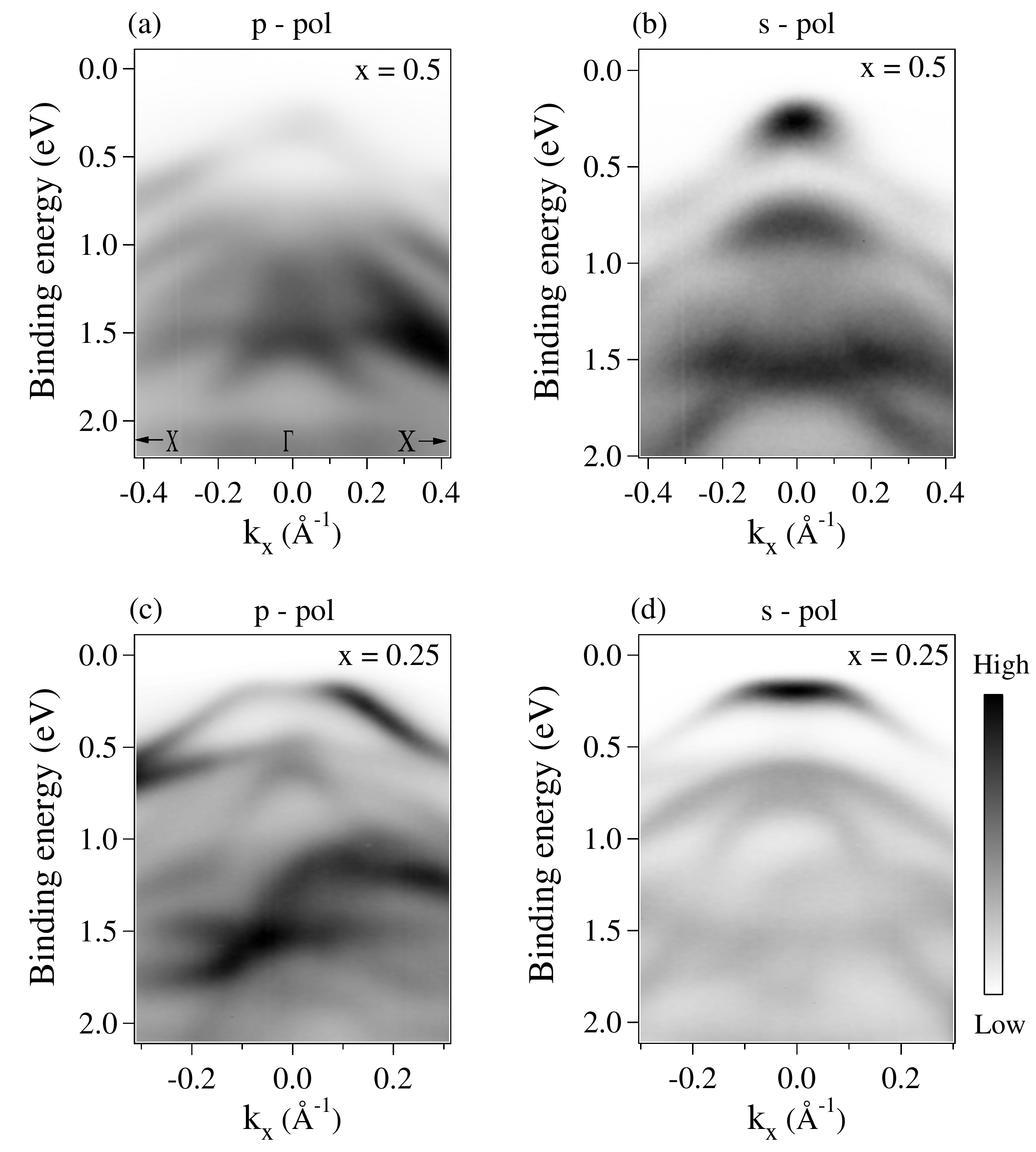}
\vspace{0ex}
\caption{ARPES intensity maps of Ta$_2$Ni(Se$_{0.5}$S$_{0.5}$)$_5$ acquired along $\Gamma-X$ direction at T = 80 K and at photon energy 26 eV using (a) p-polaried light and (b) s-polarized light. The same for Ta$_2$Ni(Se$_{0.75}$S$_{0.25}$)$_5$ in (c) and (d), respectively at photon energy 20 eV.}
\label{SuppFig1}
\vspace{-2ex}
\end{figure*}

Fig.~\ref{SuppFig1} shows the ARPES intensity plots of 25$\%$ and 50$\%$ Sulphur-doped Ta$_2$NiSe$_5$ acquired using $xz$ (p) and $y$ (s) - polarized light. The intensity of the top part of the valence band around $\Gamma$ is enhanced for s-polarization [see Fig.~\ref{SuppFig1}(b) and (d)] while the dispersive parts of the valence band away from $\Gamma$ (and towards $X$) display higher intensities for p-polarized light [see Fig.~\ref{SuppFig1}(a) and (c)]. This reveals the mirror symmetries associated with different orbitals forming the valence band in these compounds with respect to the $xz$ plane (for our experimental setup).

\clearpage

\subsection{Anisotropic in-plane band dispersions in Ta$_2$NiSe$_5$ and Ta$_2$Ni(Se$_{0.5}$S$_{0.5}$)$_5$}
\begin{figure*}[h]
\centering
\vspace{4ex}
\includegraphics[width = 0.6\linewidth]{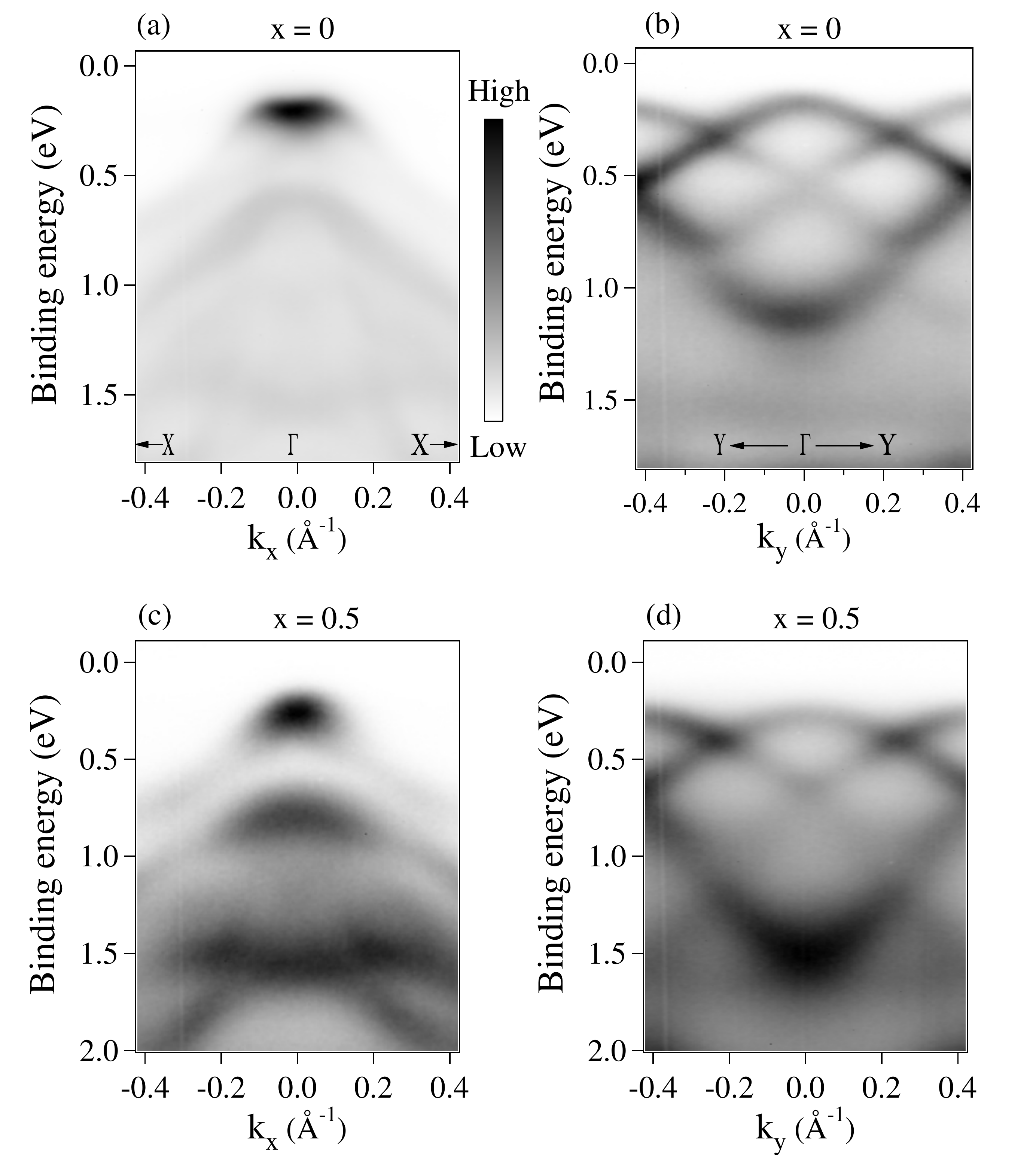}
\vspace{0ex}
\caption{(a) and (b) Anisotropic band dispersions along $\Gamma-X$ and $\Gamma-Y$ directions in Ta$_2$NiSe$_5$. Band dispersions along $\Gamma-X$ and $\Gamma-Y$ in Ta$_2$Ni(Se$_{0.5}$S$_{0.5}$)$_5$ in (c) and (d), respectively. A photon energy of 26 eV and s-polarized light is used.}
\label{SuppFig2}
\vspace{-2ex}
\end{figure*}

The in-plane anisotropy of the energy band dispersions along $\Gamma-X$ and $\Gamma-Y$ directions for Ta$_2$NiSe$_5$ having 0$\%$ and 50$\%$ of Sulphur-doping is shown in Fig.~\ref{SuppFig2}(a), (b) and Fig.~\ref{SuppFig2}(c), (d), respectively. The valence band near the Fermi level, $E_F$, exhibits a hole-like dispersion centered at $\Gamma$ and is characterized by a stronger dispersion along $\Gamma-X$ direction (parallel to the Ta-Ni-Ta chains) than along $\Gamma-Y$ direction (perpendicular to the chains). This indicates that the hopping properties of the carriers are different along the $a$-axis and $c$-axis. 

\clearpage

\subsection{Comparison of photon energy dependence of valence band peak position among Ta$_2$Ni(Se$_{1-x}$S$_x$)$_5$ compounds}
\begin{figure*}[h]
\centering
\vspace{4ex}
\includegraphics[width = 0.55\linewidth]{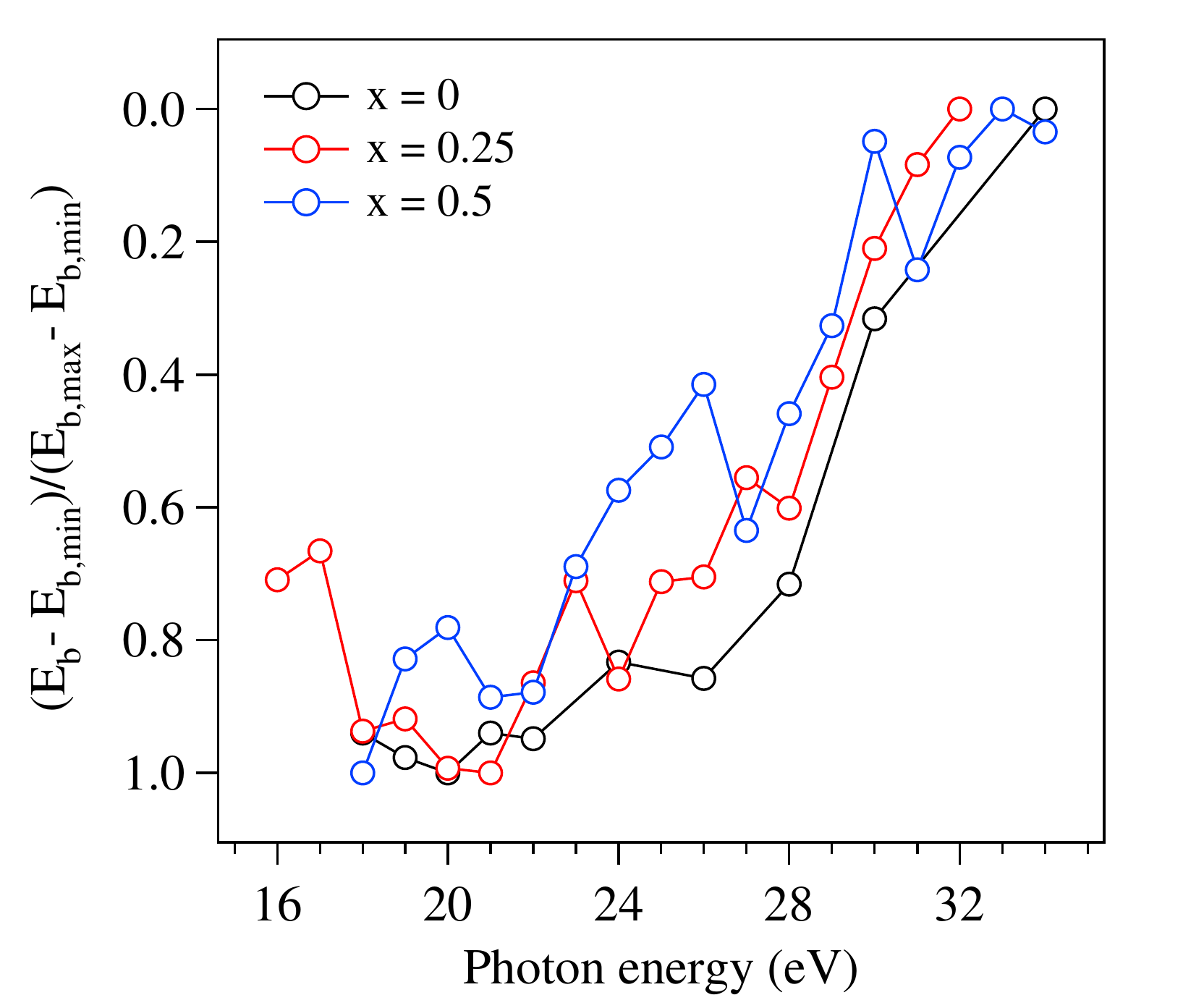}
\vspace{0ex}
\caption{Peak position of the valence band at $\Gamma$ [shown in Fig.7(g)-(i)] as a function of incident photon energy for pure and S-doped Ta$_2$NiSe$_5$, plotted on the same relative scale for comparison. E$_b$ stands for the binding energy. }
\label{SuppFig3}
\vspace{-2ex}
\end{figure*}

Fig.~\ref{SuppFig3} shows a comparison of the valence band peak position at $\Gamma$ along $\Gamma-X$ direction as a function of the incident phtoton energy for the three doped (0$\%$, 25$\%$, 50$\%$) compounds. The binding energy values have been plotted on the $(E_b-E_{b,min})/(E_{b,max}-E_{b,min})$ scale for each of the compounds. Apart from the quantitative differences in the binding energy, $E_b$, values between the three curves (red, black, blue in Fig.~\ref{SuppFig3}), the overall trend, i.e., a decrease in the binding energy with increasing photon energy, does not show any pronounced qualitative differences between the three compounds, at least within the photon energy range used for this study. \\

All the data analysis has been performed using Igor Pro (wavemetrics.com) and Casa XPS (http://www.casaxps.com) softwares and the crystal structure in Fig. 1(a) was generated in VESTA\cite{vesta}.

%\begin{figure}[t]
%\centering
%\vspace{-4ex}
%\includegraphics[width = 1\linewidth]{Figure_Suppl.png}
%\vspace{-2ex}
%\caption{(a) trARPES spectrum at a pump fluence of 3.6 mJ/cm$^2 $ around $\Gamma$ point for the integrated angular range indicated by the vertical dashed lines in Fig. 2(a) of the manusript; (b) Oscillatory part of the photoemission signal; (c) Fourier analysis of (b). }
%\label{Suppl}
%\vspace{-2ex}
%\end{figure}

%In the manuscript, we show the presence of a 2 THz phonon mode to be the dominant one, see inset of Fig. 3(b). However, an oscilllation frequency of 3 THz is observed for the energy band lying at - 0.8 eV as shown in Fig.~\ref{Suppl}. The energy integration window is shown by the horizonal dashed lines in Fig.~\ref{Suppl}(a). The oscillations in Fig.~\ref{Suppl}(b) have been extracted by subtracting an exponential decay function from the time-resolved integrated photoemission intensity in the energy interval indicated in (a) and its Fourier transform gives 3 THz as the dominant frequency mode, see Fig.~\ref{Suppl}(c).

%\bibliography{excitons,tdmft}